\documentclass[fleqn,twoside]{article}
\usepackage{espcrc2}

\usepackage{graphicx}
\usepackage[figuresright]{rotating}
\usepackage{psfrag}
\usepackage{color}
\usepackage{amssymb}

\newcommand{\ea}{\begin{eqnarray}}
\newcommand{\na}{\end{eqnarray}}

\newcommand{\Z}{{\mathbb{Z}}}
\newcommand{\N}{{\mathbb{N}}}
\newcommand{\R}{{\cal R}}
\newcommand{\U}{{\cal U}}
\newcommand{\La}{{\cal L}}
\newcommand{\eq}{\begin{equation}}
\newcommand{\en}{\end{equation}}
\newcommand{\um}{\frac12}

\newcommand{\bra}{\langle}
\newcommand{\ket}{\rangle}
\newcommand{\oline}{$\bullet$---\hskip -1pt---${\hskip -2pt\bullet}$}

\newcommand{\ww}{\mbox{
\begin{minipage}{16pt}\includegraphics[width=16pt]{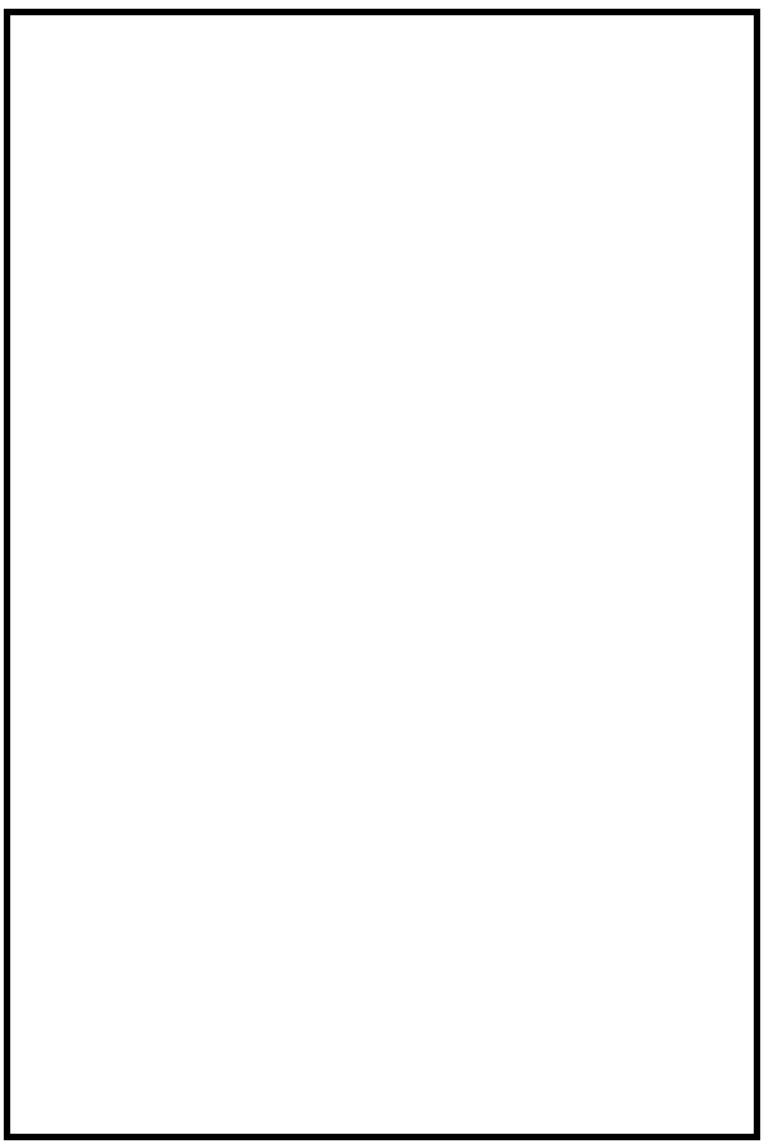}\end{minipage}}}

\newcommand{\wbbd}{\mbox{
\begin{minipage}{16pt}\includegraphics[width=16pt]{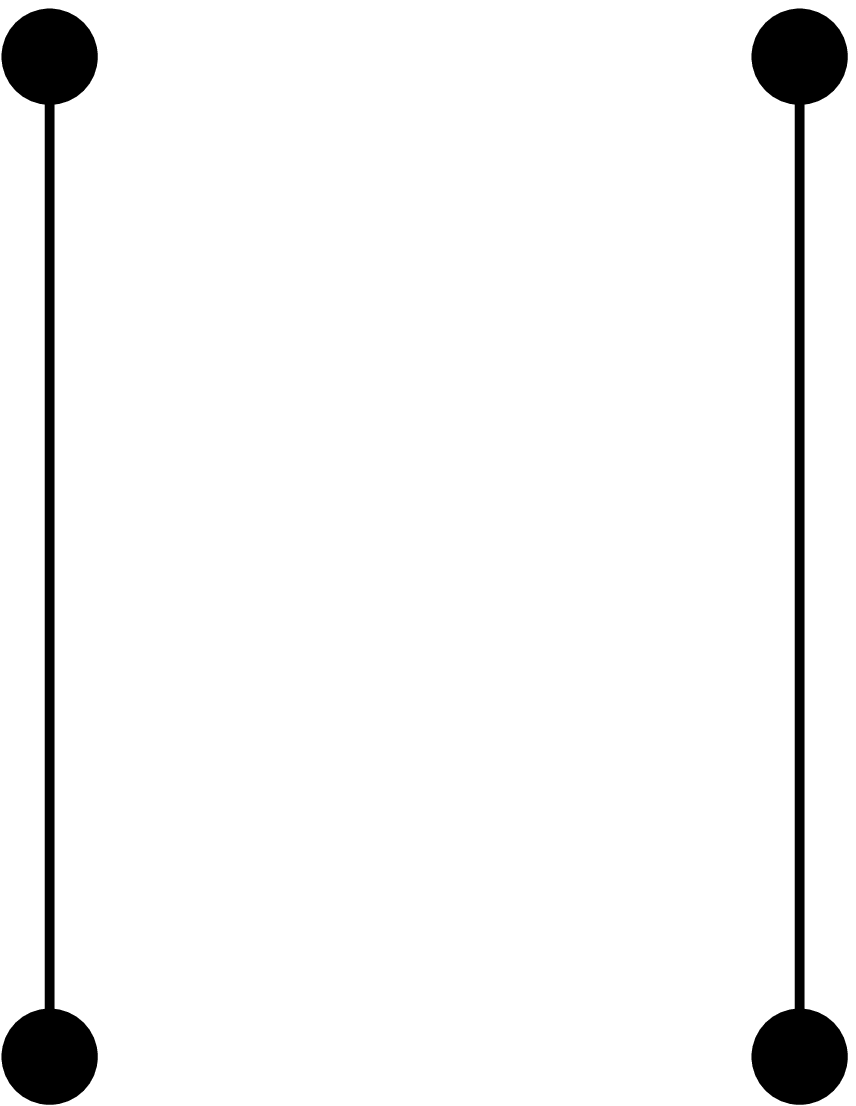}\end{minipage}}}
\newcommand{\wwb}{\mbox{
\begin{minipage}{16pt}\includegraphics[width=16pt]{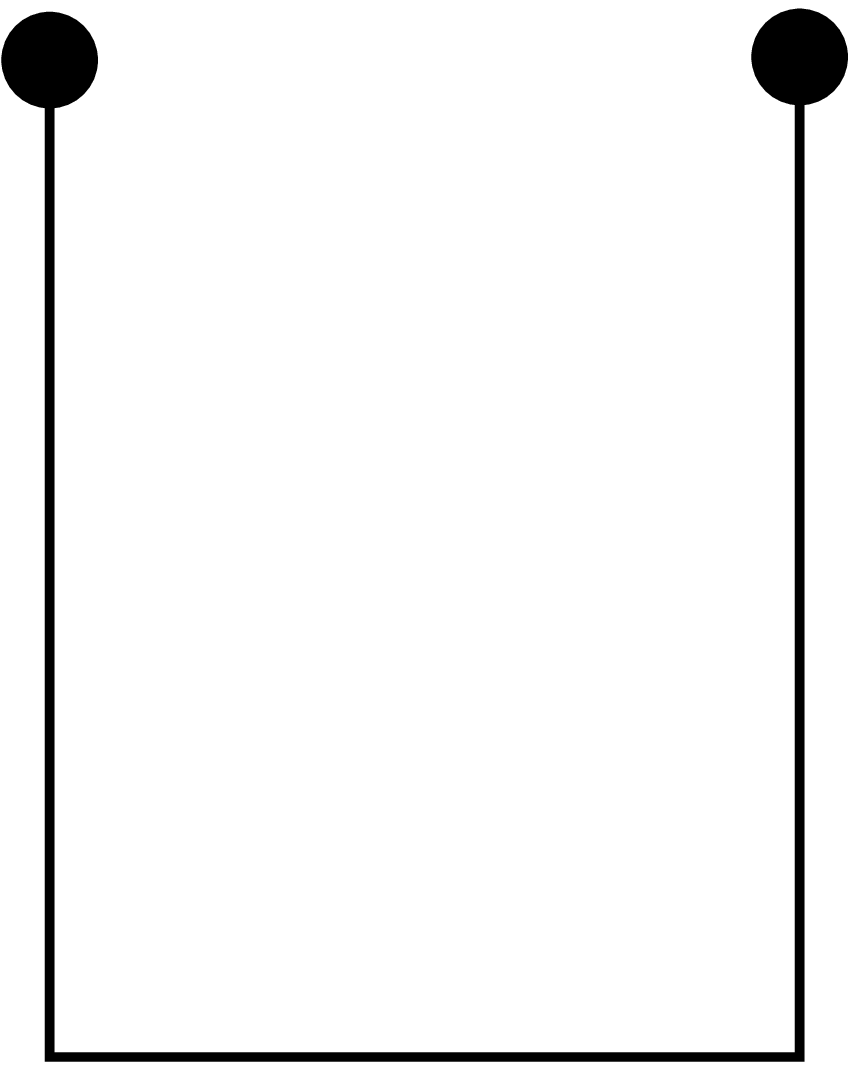}\end{minipage}}}
\newcommand{\wbw}{\mbox{
\begin{minipage}{16pt}\includegraphics[width=16pt]{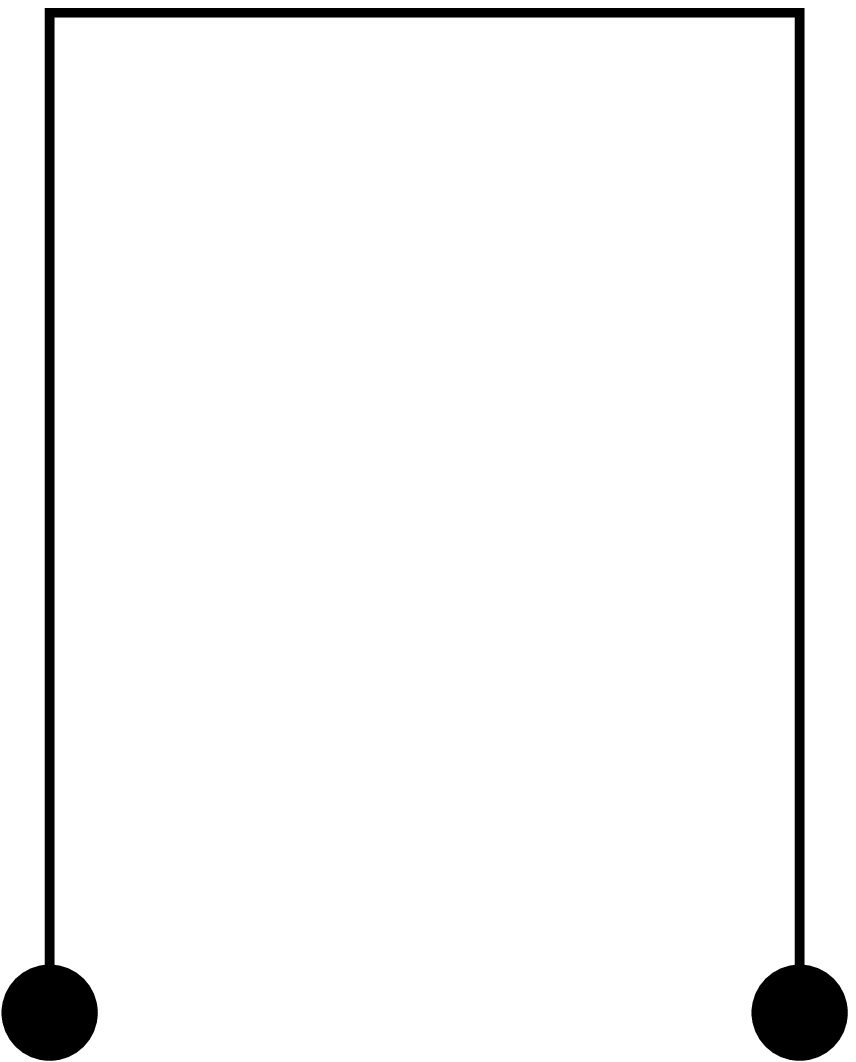}\end{minipage}}}
\newcommand{\uu}{\mbox{
\begin{minipage}{10pt}\includegraphics[width=10pt]{pictures/u12.eps}\end{minipage}}}
\newcommand{\uleft}{\mbox{
\begin{minipage}{12pt}\includegraphics[angle=90,width=12pt]
{pictures/u12.eps}\end{minipage}}}

\newcommand{\uright}{\mbox{
\begin{minipage}{12pt}\includegraphics[angle=-90,width=12pt]
{pictures/u12.eps}\end{minipage}}}

\newcommand{\ulright}{\mbox{
\begin{minipage}{26pt}\includegraphics[angle=-90,width=26pt]
{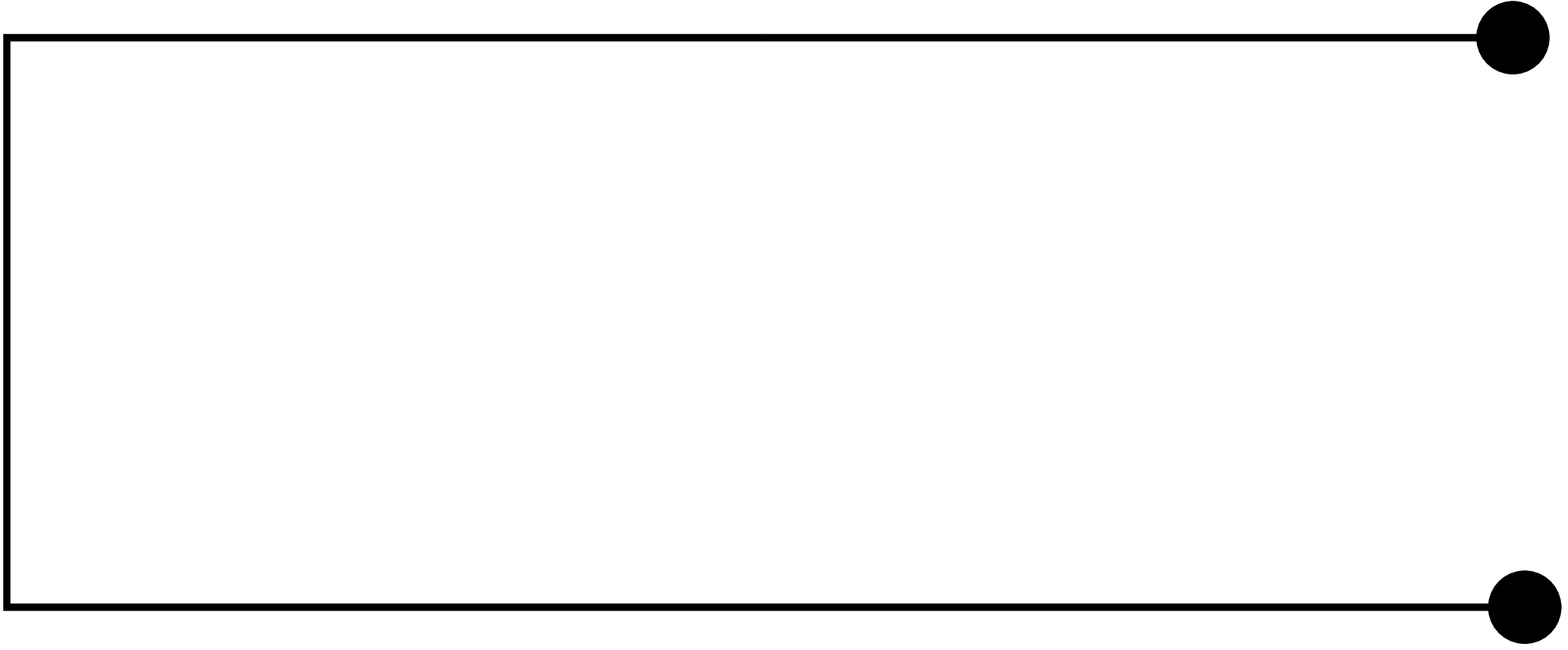}\end{minipage}}}

\newcommand{\voline}{\mbox{
\begin{minipage}{1.2pt}\includegraphics[width=1.2 pt]{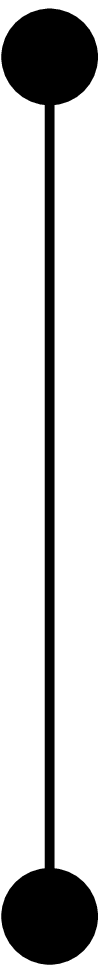}\end{minipage}}}

\newcommand{\displa}{\mbox{\begin{minipage}{6 cm}
\includegraphics[width=6 cm]{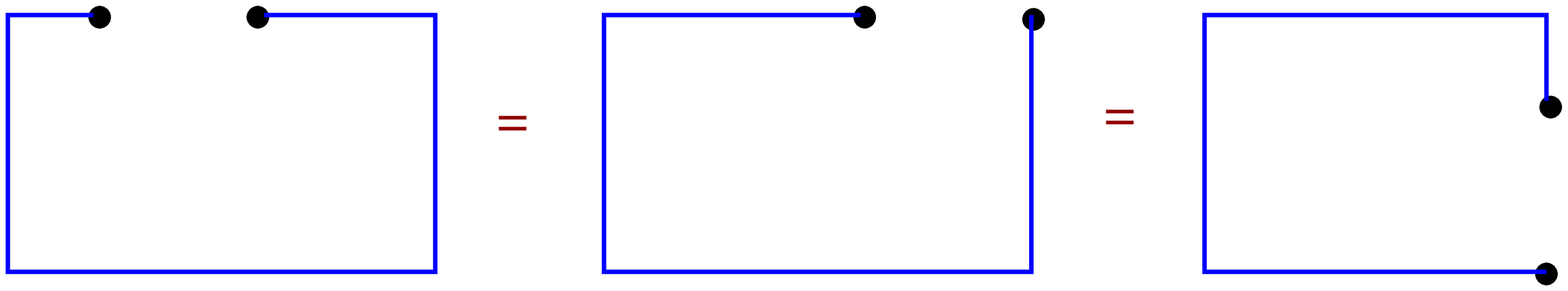}\end{minipage}}}
\newcommand{\diag}{\mbox{\begin{minipage}{6 pt}
\includegraphics[width=6 pt]{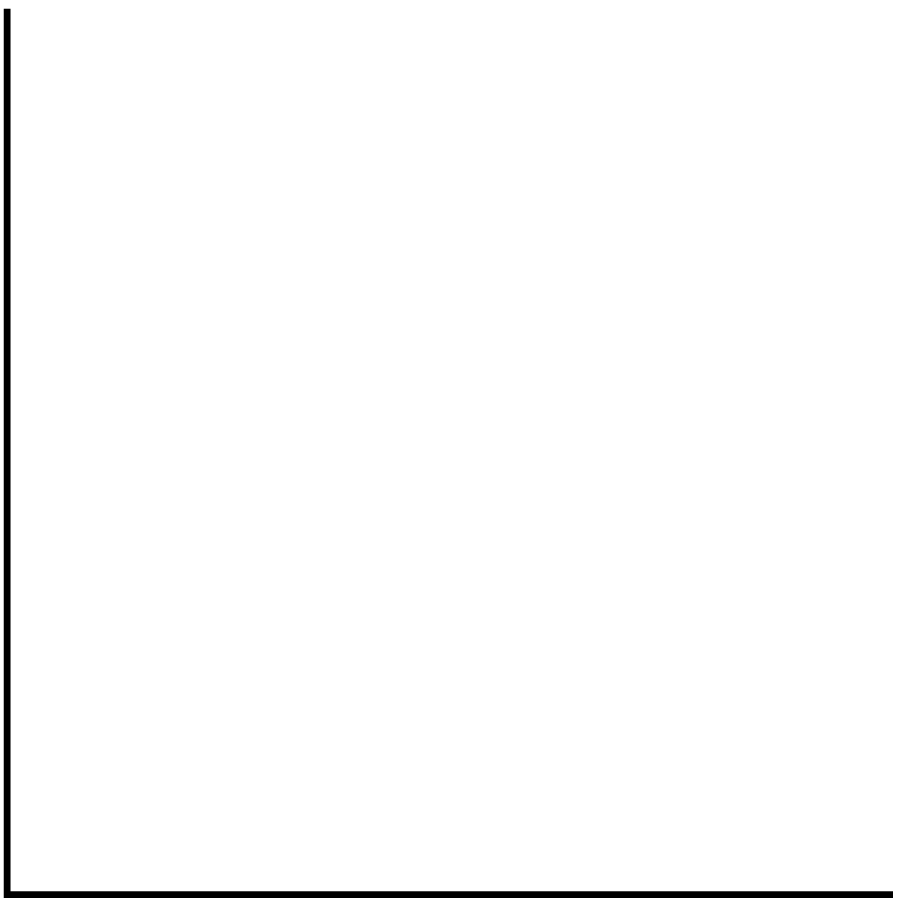}\end{minipage}}}

\newcommand{\AmS}{{\protect\the\textfont2
  A\kern-.1667em\lower.5ex\hbox{M}\kern-.125emS}}

\title{The functional form of open Wilson lines \\  
 in gauge theories coupled to matter}

\author{F. Gliozzi\address{Dipartimento di Fisica Teorica, 
Universit\`a di Torino,\\ 
       INFN, Sezione di Torino, via P. Giuria 1, 10125 Torino, Italy}}

\begin{document}

\begin{abstract}
The open Wilson lines are gauge-invariant operators made with a 
gauge transporter along an open path  
saturated at the end-points with matter fields. Here it is shown that 
numerical experiments on 3D $\Z_2$ Higgs model provide useful guidance in 
addressing the problem of the functional form of their vacuum 
expectation values. It turns out they satisfy, as long as their size 
does not exceed the string breaking scale, a remarkable 
factorization, related to finite-size scaling of the 
Fredenhagen-Marcu order parameter. This leads to conjecture 
a decay area law, like  ordinary Wilson loops, with the 
difference that the boundary conditions of the confining string world-sheet 
are fixed along the path of the gauge transporter whereas are free 
on the straight line joining the two end-points. A consistency condition 
fixes uniquely the contribution of the free boundary, which turns out 
to be proportional to the string-breaking scale. Universal shape effects 
produced by quantum string fluctuations are also studied.       
\end{abstract}

\maketitle

\section{INTRODUCTION}
The most basic observables of gauge theories are the Wilson loops. These are operators 
$W_\gamma$ associated to any closed path $\gamma$ of the space-time. Their importance 
stems from the fact that they serve as order parameters of confinement. The confining phase 
shows up in an 
\emph{area law} for the vacuum expectation value  of large Wilson loops: if $\gamma$ is scaled 
up keeping its shape fixed and increasing the encircled minimal area $A$, then 
$\bra W_\gamma\ket$   vanishes exponentially with $A$: 
$\bra W_\gamma\ket\propto e^{-\sigma\,A}$, where the physical quantity 
$\sigma$ can be interpreted as the tension of the 
confining string where the chromoelectric flux is concentrated. 
When dynamical matter is added to such a system this string breaks, 
because pairs of matter particles form 
and act as new end points of the confining string. The broken 
string state describes a bound 
state of a static colour source -the fixed end of the string- and a 
dynamical matter field 
-the free end of the string.

Contrary to earlier expectations, the string breaks only when its length 
exceeds 
a certain threshold value $\R_b$. The reason, as we shall see shortly, is 
very simple.
At intermediate distances the behaviour of the Wilson loop is indistinguishable
from that of the confining phase of the pure gauge theory. Nonetheless, 
when the Wilson loop is very large it should decay with a perimeter law, 
hence one can consequently assume that
\eq
\bra W_\gamma\ket\simeq c_u\,e^{-\sigma\,A_\gamma-\lambda\vert\gamma\vert }
+c_b\,e^{-\mu\vert\gamma\vert}
\label{ansazw}
\en
where $\vert\gamma\vert$ is the length of $\gamma$. The first term, corresponding 
to the unbroken string regime, should be implemented with a shape-dependent factor 
due to quantum fluctuations of the confining 
string \cite{Luscher:1980fr}. Such a universal shape effect has been observed also in a 
gauge theory coupled to matter \cite{Gliozzi:2000yg} as a 
further support to the conjecture 
that the confining string in this more general medium behaves exactly like the 
string in the pure gauge vacuum as long as it is shorter than $\R_b$ .
Eq.(\ref{ansazw}) can be used to describe the main features of the string 
breaking mechanism as a level crossing phenomenon. In particular the string 
breaking scale turns out to be
\eq
\R_b\simeq2\,\frac{\mu-\lambda}{\sigma}~~,
\label{rb}
\en
as expected \cite{Bock:1988kq}.
Notice that $\mu$ and $\lambda$, being self-energy contributions of the static sources, 
are UV divergent, however these divergences cancel in their difference, hence 
$\R_b$ is a meaningful physical scale even in the continuum limit.

\begin{figure}[htb]
{\psfrag{+z}{$3D~\Z_2$ Higgs model}
\psfrag{gr}{$~$}
\psfrag{hep}{$~$}
\psfrag{R/a}{$r$}
\psfrag{W(R,T=15a)}{$W(r,t=15a)$}
\centering{\includegraphics[width=7.6 cm]{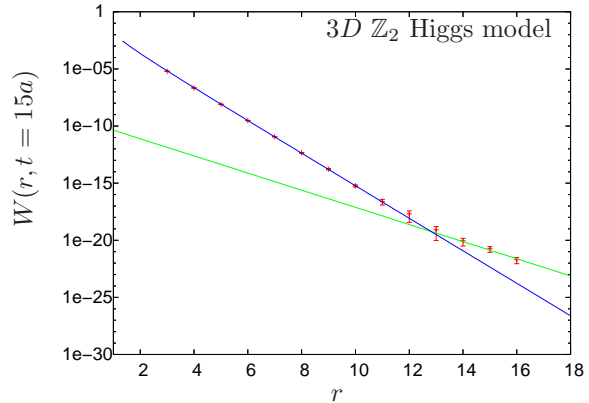}}
}
\caption{Observing string breaking with a Wilson loop in 3D $\Z_2$ Higgs model.
The steepest line is the area term of Eq.(\ref{ansazw}); the other is the perimeter
term.}
\label{Figure:0}
\end{figure}

From a computational point of view it is very hard to observe string breaking 
directly through the estimate of large Wilson loops, because Eq.(\ref{ansazw})
implies \cite{Gliozzi:2004cs,Gliozzi2005:dv} that in the loops of 
\emph{finite} size such a phenomenon 
is visible at distances much larger than $\R_b$, where the signal is 
drowned in the noise. As a matter of fact direct observation of string breaking 
by fitting the numerical data to Eq.(\ref{ansazw}) as been reported only in two 
3d systems: the SU(2) gauge theory with adjoint sources \cite{Kratochvila:2003zj} 
and the $\Z_2$ Higgs model \cite{Gliozzi:2004cs} (see Fig.\ref{Figure:0}).   

In gauge theories coupled to matter one can construct gauge invariant operators
that are more general than the Wilson loops. They are formed by a gauge
transporter along an open path $\gamma$, saturated at the end points by the 
matter fields. We call these operators  \emph{open Wilson lines}. An important example 
is the U-shaped operator $\U(r,t)\equiv$ \uu, where the black dots represent the 
matter fields. 
It can be considered as an operator connecting the string state {\bf ---} of length $r$ 
 at the bottom $t=0$ with the broken string state $\bullet~~~\bullet$ at the top,
hence it is expected to have a large overlap with the broken string state.
This suggested enlarging the basis of operators to extract the static potential
by considering the correlation matrix 

\eq
C(r,t)=\left(\begin{array}{cc}
 \ww&\quad \wwb\\&\\
\wbw&\quad \wbbd\end{array}\right),\label{eq:sb}
\label{cmatrix}
\en
 In this way a rather abrupt crossover between 
string-like and broken string states has been clearly seen at the expected 
distance $\R_b$.
So far string breaking has been verified in various theories coupled 
 to a fundamental scalar field 
 in 2+1 \cite{Philipsen:1998de} and 3+1 dimensions \cite{Knechtli:1998gf}, 
in QCD with two flavours \cite{Bali:2005fu}, as well as breaking 
of the adjoint string in 2+1 \cite{Stephenson:1999kh} and 3+1
dimensions \cite{deForcrand:1999kr}. 

It is natural to ask what is the functional form of open Wilson lines which generalises
Eq.(\ref{ansazw}). Experience suggests that the answer to this question should not 
depend on the gauge group nor on the type of  dynamical matter. Thus I performed an 
exploratory study on 
a particularly simple system, the $\Z_2$ Higgs model in three dimensions, probing open Wilson lines of various shapes. It turns out they fulfil remarkable factorization 
properties (see Eq.s (\ref{fr}) and (\ref{proposal}) below) They  lead to conjecture a 
generalisation of Eq.(\ref{ansazw}) which can be applied to any large 
\emph{planar open path} $\gamma$ joining two points placed at a distance $r$
\begin{eqnarray}
\nonumber
\bra G_r(\gamma)\ket\simeq c_l \,e^{-\sigma\, A_{\widetilde\gamma}
-\lambda\vert\gamma\vert-\R_b\,\sigma\, r/2}+&\\
\rho\, c_l\,e^{-\sigma\, A_{\widetilde\gamma}
-\lambda\vert\gamma\vert +\R_b\,\sigma\, r/2}+&\cdot\cdot\cdot
\label{genera}
\end{eqnarray}

\begin{figure}[htb]
{
\psfrag{R}{$r$}
\psfrag{G}{$\gamma$}
\psfrag{A}{$A_{\widetilde{\gamma}}$}
\centering{\includegraphics[width=6.5cm]{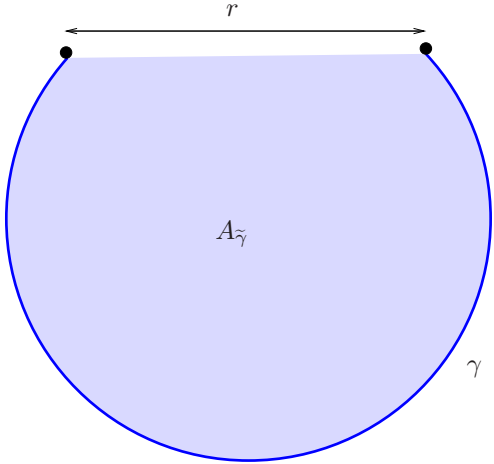}}
}
\caption{Open Wilson line. The solid dots represent the matter fields; the shaded 
region denotes the world-sheet of the underlying confining string.}
\label{Figure:1}
\end{figure}

where $\widetilde{\gamma}=\gamma+\ell$ is the closed path  obtained by joining the 
two end points of  
$\gamma$ with the straight line $\ell$, $A_{\tilde\gamma}$ is the area of the planar 
surface it encircles, $\rho$ is (a generalisation of) the Fredenhagen-Marcu order 
parameter \cite{Fredenhagen:1985ft},
  and the ellipses denote sub-dominant terms at intermediate distances 
which become visible only at large scale, where the asymptotic behaviour is
 dominated by the perimeter law, like in Eq.(\ref{ansazw}). 
The first term of Eq.(\ref{genera}) is similar to the first term of an older proposal 
\cite{Bock:1988kq} with the important specification that the coefficient of $r$ in the 
exponent, which can be regarded as the self-energy contribution of the constituent 
Higgs particle, is here directly related to the string breaking scale. 
When 
$\gamma=\ell$ then $A_{\widetilde\gamma}=0$ and  we get the straight line operator 
\oline$~=G_r(\ell)\equiv\La_r$. In such a case  
Eq.(\ref{genera}) is replaced  by the following asymptotic expansion
\eq
\bra \La_r\ket \simeq c_l\, e^{-\mu\,r}+\,\cdot\cdot\cdot
\label{oline}
\en
where $\mu$ is, like in Eq.(\ref{ansazw}), the mass of the ground state of the so called 
static-light meson. Eq.(\ref{genera}) can be 
still interpreted in terms of the underlying confining string  
(see Fig.\ref{Figure:1}). Clearly this has 
fixed boundary conditons only along $\gamma$, while it is free to vibrate 
along $\ell$. Also in this case we expect that the area term should be modified  
by universal shape effects due to quantum fluctuations of the string. An explicit 
form of such a contribution in the case of the U-shaped operator is calculated in 
\S~\ref{shape}.
\section{ALMOST CLOSED WILSON LINES}
The $\Z_2$ Higgs model which has been analysed in detail by a number of
Monte Carlo simulations is defined through the action
\eq
S==-\beta_G\sum_PU_P-\beta_I\sum_x\sum_{\mu=1}^3\phi_xU_{x,\hat\mu}
\phi_{x+\hat\mu}~.
\label{action}
\en
Here $U_{x,\hat\mu}$ and $U_P$ are link and plaquette variables respectively, which describe the gauge degrees of freedom; the site variables $\phi_x$ represent the matter fields. 
Both $U_{x,\hat\mu}$
and $\phi_x$  take values on the $\Z_2$ gauge group.  This model is self-dual and has a 
confinement/Higgs phase and a Coulomb-like phase \cite{Jongeward:1980wx}. All 
the MC simulations are performed in the region  $0<\beta_G<0.755$, $0<\beta_I<0.248$ 
which is called confinement ``phase'', because the Wilson loops of intermediate size 
decay with an area law.  

\begin{figure}[htb]
{\psfrag{W}{\small Square~ Wilson ~Loop}
\psfrag{U}{\small U-shaped Operator}
\psfrag{L}{$l$}
\includegraphics[width=7.5cm]{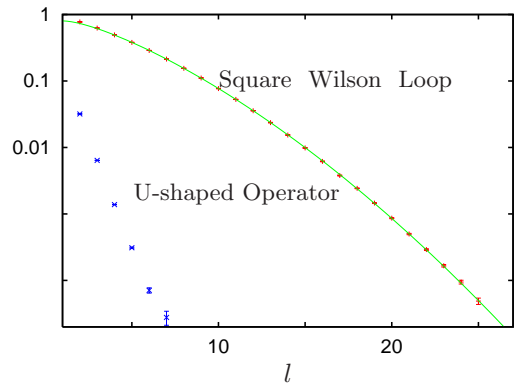}}
\caption{A comparison between square Wilson loops and U-shaped operators of the 
same size at $(\beta_G,\beta_I)=(0.75245,0.16683)$ where 
$\R_b\,\sqrt{\sigma}\sim27$.}
\label{Figure:2}
\end{figure}
Measuring $\bra W(r,t)\ket$ and $\bra \La_r\ket$  yields an estimate of the 
string breaking scale $\R_b$. While 
$\bra W\ket$, as it turns out, does not depend on  $\R_b$ at the probed scales, 
the U-shaped operator is very sensitive to it: $\bra\U(r,t)\ket$ 
drops off rapidly as $\R_b$ increases (see Fig.\ref{Figure:2}). This is not 
unexpected: when $\R_b\to\infty$ the theory becomes a pure gauge theory, therefore
$\bra \U\ket\equiv0$. 

\begin{figure}
{\psfrag{R}{$r$}
\psfrag{T}{$t$}
\psfrag{L}{$l$}
\includegraphics[width=6.5 cm]{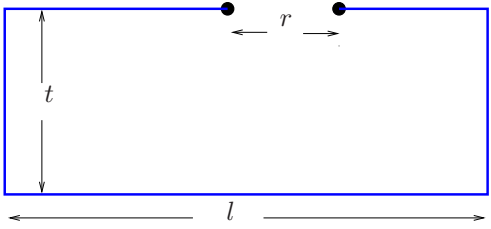}
}
\caption{The almost closed Wilson lines interpolate between the Wilson loops and 
the U-shaped operators.}
\label{Figure:3}
\end{figure}

In order to investigate in more detail such a dependence on the string breaking scale 
it is convenient to study a new family of gauge-invariant operators, that we call 
\emph{almost closed Wilson lines} $G_r(l,t)$, which somehow interpolate 
between the U-shaped ones and the Wilson loops (see Fig.\ref{Figure:3}).
It turns out that the vacuum expectation value (vev) of these operators does not depend very 
much on the position of the pair of matter sources $\bullet~~\bullet$  along $\gamma$. 
This can be expressed with the graphical equation 
\eq
{\psfrag{=}{$\simeq$}
\displa
}~~~.
\label{displa}
\en

\begin{figure}[htb]
\psfrag{R=1}{\textcolor{blue}{ $ r=1$}}
\psfrag{R=2}{\textcolor{blue}{ ${ r=2}$}}
\psfrag{R=3}{\textcolor{blue} {${ r=3}$}}
\psfrag{R=4}{\textcolor{blue} {${ r=4}$}}
\psfrag{R=5}{\textcolor{blue} {${ r=5}$}}
\psfrag{R=6}{\textcolor{blue} {${ r=6}$}}
\psfrag{R=7}{\textcolor{blue} {${ r=7}$}}
\psfrag{L}{$l$}
\psfrag{title}{\textcolor{black}
{ $\bra W(l,l)\ket/\bra G_r(l,l)\ket $}}
\mbox{\includegraphics[width=7.7cm]{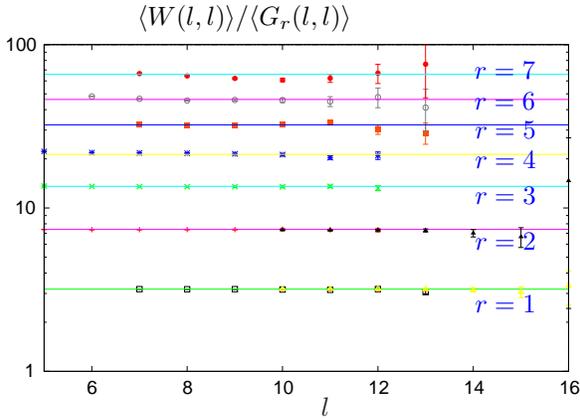}}
\caption{Ratio between a square Wilson loop $\bra W(l,l)\ket$ and an 
almost closed square Wilson line of the same size $\bra G_r(l,l)\ket$. One 
clearly sees that such a ratio does not depend on the 
size $l$, but only on the distance $r$ between the end points.} 
\label{Figure:4}
\end{figure}
\begin{figure}[bht]
{\psfrag{R}{$r$}
\psfrag{H(R)}{$f(r)$}
\centering{\includegraphics[width=8.0cm]{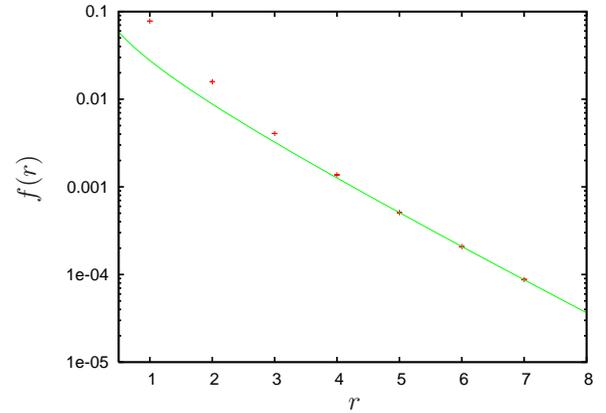}}
}
\caption{The function $f(r)$ defined in Eq.(\ref{fr}) evaluated at
$(\beta_G,\beta_I)=(0.735,0.224)$. The continuous line is a fit to
$\frac{c}re^{-m\,r}$. The parameters $c$ and $m$ will be related to other 
physical quantities of the system.}
\label{Figure:5}
\end{figure}

 Evaluating the ratio $\bra W(l,l)\ket/\bra G_r(l,l)\ket$ 
one discovers the surprising result that it does not depend, within the errors,
on the size $l$, as Fig.\ref{Figure:4} demonstrates. Although 
this numerical fact cannot directly promoted to the rank of a physical property 
of the continuum theory because of unbalanced self-energy divergences 
between numerator and denominator, by combining the almost closed Wilson 
line $G_r(l,l)$ with the straight line operator $\La_r$ one is led to consider the 
following gauge-invariant, UV finite  quantity  
\eq
\frac{\bra\La_r\ket\,\;\bra G_r(l,l)\ket}{\bra W(l,l)\ket}\simeq f(r)~~,
\label{fr}
\en
which, according to the results of Fig.\ref{Figure:4}, seems to depend only on $r$. The 
behaviour of such a function is drawn in Fig.\ref{Figure:5}. More insight into $f(r)$ 
will be gained in the next section. 

\section{FACTORIZATION}
\label{factorization}
There is no reason to believe that the almost closed Wilson lines play some special 
role in gauge theories, thus Eq.(\ref{fr}) should be regarded as reflecting a 
more general property of whatever open Wilson line. 
\begin{figure}
{\psfrag{a}{$\alpha$}
\psfrag{g}{$\gamma$}
\psfrag{b}{$\beta$}
\psfrag{>}{$\to$}
\includegraphics[width=7. cm]{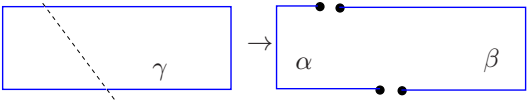}
}
\caption{Splitting a Wilson loop into two open Wilson lines.}
\label{Figure:6}
\end{figure}
An obvious generalisation which comes to mind is to consider an arbitrary 
planar Wilson loop $W(\gamma)$ (see Fig.\ref{Figure:6}), cut the closed 
path $\gamma$ in two 
points  placed at a distance $r$ and  saturate the end points of the resulting
open paths $\alpha$ and $\beta$ with matter fields. In this way one is led to 
write down the following UV finite quantity, generalising Eq.(\ref{fr})
\eq
\frac{\bra G_r(\alpha)\ket\;\bra G_r(\beta)\ket}
{\bra W(\gamma=\alpha+\beta)\ket}\;\simeq\;f(r)~~.
\label{proposal}
\en
In order to lend some support to such a proposal we measured the ratios 
$\bra \La_r\ket\,\bra\U(r,r)\ket/\bra W(r,r)\ket$ , 
$\bra\U(r,r)\ket^2/\bra W(2r,r)\ket$ and 
$\bra G_{\sqrt{2}\,r}(\diag)\,\ket^2/\bra W(r,r)\ket$, where $\diag$ denotes the path 
obtained by cutting a square in half along its diagonal.
\begin{figure}[htb]
{\psfrag{R}{$r$}
\mbox{\includegraphics[ width=8 cm]{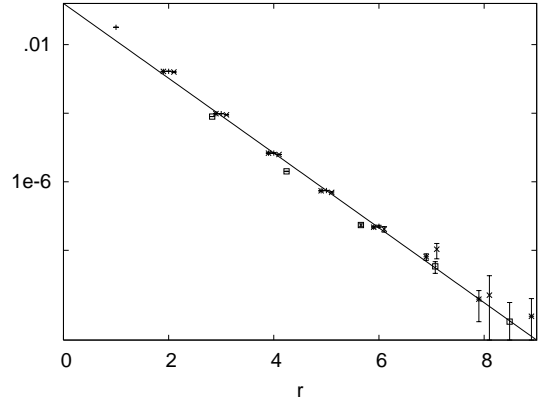}}
}
\caption{Testing Eq.(\ref{proposal}). The $\Box$'s correspond to
$\bra G_{\sqrt{2}\,r}(\angle)\,\ket^2$, the $\star$'s correspond to $\bra\U(r,r)\ket^2$
and the $\times$'s to $\bra\La_r\ket\,\bra \U(r,r)\ket$, finally the $+$'s and the 
continuous line have the same meaning as in  Fig.\ref{Figure:5}, but here the data are 
taken at $(\beta_G,\beta_I)=(0.75245,0.16683)$. Data corresponding to the same 
value of $r$ are displaced for clarity.}   
\label{Figure:7}
\end{figure}
Some results are presented in Fig.\ref{Figure:7}. 

Though these numerical tests of
Eq.(\ref{proposal}) are manifestly very good, there are regions of the parameter 
space where the agreement is less good. The general rule seems to be that 
Eq.(\ref{proposal}) is fulfilled, at least approximately, whenever the size of the involved loop is less than the string breaking scale. But even in this region 
Eq.(\ref{proposal}) cannot reflect an exact property of the gauge theories. 
The reason is very simple: the vev of the Wilson loop 
$W(\gamma)$  in the denominator depends explicitly on the  shape of $\gamma$ and this 
dependence cannot be counterbalanced by the shape effects of the numerator, 
being the splitting $\gamma\to \alpha+\beta$  completely arbitrary. Such a fact will 
be strengthened in the next section by an explicit calculation of the universal 
shape effects generated by the quantum fluctuations of the underlying confining 
string in the U-shaped open Wilson line. 

In conclusion, Eq.(\ref{proposal}) has no chance to be exact.  Rather, it should be 
viewed as the first term of an expansion in some scale parameter, perhaps 
$1/\sigma A_\gamma$. Note that if $\gamma$ is scaled up to $\infty$ keeping its 
shape fixed one gets (a generalisation of) the Fredenhagen-Marcu order parameter  
 \cite{Fredenhagen:1985ft}  
\eq
\lim_{r\to\infty}{~}\frac{\bra G_r(\alpha)\ket\;
\bra G_r(\beta)\ket}
{\bra W(\gamma)\ket}{~}_{\begin{array}{|c}
{\rm fixed}\\
{\rm shape}
\end{array}}  =\rho ~~,  
\label{rho}
\en
which vanishes in the Coulomb-like phase while it is different from zero in the 
confinement/Higgs phase.

Eq.(\ref{proposal}) suggests the first correction to $\rho$ should be a 
shape-independent
term. Being only a function of the distance $r$ of the two end points where are sitting the matter fields, it is conceivable that $f(r)$ has the asymptotic expansion of a two-point correlation function, namely
\eq
f(r)=\rho+\frac c{r^{d-2}}e^{-m\,r}+\frac{c'}{r^{d-2}}e^{-m'r}+\cdot\cdot\cdot~,
\label{asyf}
\en
which fits well to the numerical data as Fig.\ref{Figure:5} and \ref{Figure:7} show.
The parameter $\rho$ is too small and can be put safely equal to zero in these fits. 
A better evaluation of $\rho$ will be obtained shortly.

It has to be noted that the ``mass'' $m$, though is a physical quantity, 
needs not to correspond to the mass of any state of the physical spectrum, 
being associated not to a single correlator, but to a 
suitable rational function of three of them. A previously 
unsuspected  relation between $m$ and the string breaking scale will be 
uncovered below. 

Neglecting for simplicity the pre-exponential factor $1/x^{d-2}$ and using as 
input the Ansatz (\ref{oline}) we get
\eq
\bra G_r(\gamma)\ket\simeq \bra W(\widetilde\gamma)\ket( c\,e^{(\mu-m)\,r}+\rho)/c_l~.
\label{grw}
\en
The first term shows that open Wilson lines obey a perimeter-area law which 
differs from that of the ordinary Wilson loops  in two respects. First, the 
proportionality constant is different. Secondly, there is no analogue of 
the contribution of that part of the boundary 
where the string is free, encoded in the coefficient $c_f$ of $r$ in the exponential, 
which  clearly is $c_f=\mu -\lambda-m$. 

In order to find the link between $m$ and $\R_b$ we alluded to above, the clue is provided 
by the observation that Eq.(\ref{proposal}) implies the consistency condition
\eq
\bra \U(r,t)\ket^2\simeq\bra\U(r,2t)\ket\;\bra\La_r\ket
\en
or, diagrammatically,
\eq
\bra\uright\,\ket\;\bra\uleft\,\ket\simeq \bra\ulright\,\ket \,
\bra\voline\,\ket~.
\label{consi}
\en
It gives at once
\eq
c=\frac{c_l^2}{c_u}~;~c_f=-\frac m2~~;~m=\sigma\R_b~~~.
\label{cost}
\en
Combining Eq.s (\ref{grw}) and (\ref{cost}) we get 
finally Eq.(\ref{genera}), which was the major purpose of this talk.

In a nutshell, we may conclude that  the surprising factorization properties
of the vev of the open Wilson lines tell us nothing more than their behaviour
at intermediate scales is dominated by the exponential law (\ref{genera}).

One could try to look for a more detailed  solution of (\ref{proposal}) 
by taking into account the pre-exponential factor $1/r^{d-2}$ of 
Eq.(\ref{asyf}). Note  however that such a kind of correction is of the 
same order of the already mentioned shape effects, which we know do not obey 
factorization, thus  using factorization to find 
the sub-dominant corrections to the exponential behaviour is no longer 
justified. 

\begin{figure}
{\psfrag{R}{$r$}
\psfrag{U}{$\U(r)$}
\includegraphics[width=8. cm]{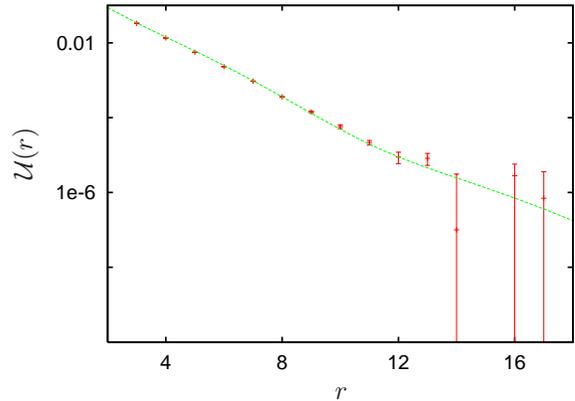}
}
\caption{The U-shaped Wilson line at $(\beta_G,\beta_I)=(0.735,0.224)$. 
The change in slope is due to the non-vanishing of $\rho$, which is the only 
free parameter of the fit.}
\label{Figure:8}
\end{figure}

An interesting property of the first two terms of Eq.(\ref{genera}) is that 
all the parameters but one (the Fredenhagen- Marcu order parameter) can be 
determined by evaluating the  vev of ordinary Wilson loops or  straight 
Wilson lines, hence the fits to this equations are one-parameter fits. As 
an example a fit to the U-shaped Wilson line is presented in 
Fig.\ref{Figure:8}. Note that the dependence on $\rho$ is strongly enhanced 
by an exponentially increasing factor.

\section{UNIVERSAL SHAPE EFFECTS}
\label{shape}
It is known that the description of the vev of   Wilson loops  as a simple 
exponential like in (\ref{ansazw}) is not sufficiently accurate.
Even in gauge theories coupled with matter the confining string, as long as 
its length does not exceed $\R_b$, can vibrate freely exactly like in 
the absence of dynamical matter \cite{Gliozzi:2000yg,Gliozzi:2004cs}. 
It is then natural to
conjecture that also in the open Wilson line $G_r(\alpha)$ one could 
detect universal shape effects generated by the quantum fluctuations of 
the confining string like in the ordinary Wilson loop. The  important 
difference is that now the boundary conditions of the string are  
fixed only along the open path $\alpha$, whereas are free
in the part of the boundary formed by the straight line connecting its 
end-points. 

In the case 
of the U-shaped operator it is easy to 
evaluate the contribution of these fluctuations. 
To be definite, consider the operator $\U(r,t)$, corresponding to the case 
$r=l$  of Fig.\ref{Figure:3}. The boundary conditions are chosen accordingly. 
The universal shape effects are encoded in the quantity 
\cite{Luscher:1980fr,aop}
\eq
\left[{\rm det}(-\partial^2)\right]^{-\frac{d-2}2}
\equiv \vert-\partial^2\vert^{-\frac{d-2}2} ~,
\label{det}
\en
where $\partial^2$ is the Laplacian of the 2d bosonic field 
describing the 
string displacements in the $d-2$ transverse directions with respect to the minimal 
surface. We have
\eq
\vert-\partial^2\vert=\prod_{m>0}\prod_{n>0}\frac{\pi^2}{r^2}
[n^2+\frac{r^2}{t^2}(m-\um)^2]~,
\label{lapla}
\en
which clearly needs regularization. A simple method
\cite{me} is based on the 
assumption that regularised series (or products) fulfil some formal 
properties of the absolutely convergent series. In particular, we assume
the two relationships
\ea
\sum_{n>0}'n^k=2^k[\sum_{n>0}'n^k+\sum_{n>0}'(n-\um)^k],&
\label{suma}\\
\sum_{n>0}'n^k=\sum_{n=1}^{a-1}  n^k+\sum_{n>0}'(n+a)^k,&
\label{sumb}
\na
where the apex indicates the regularised expression and the finite sum 
$ \sum_{n=1}^{a-1}= P_{k+1}(a)$ is a polynomial of degree $k+1$. For our 
purpose we need only to know $P_1(a)=a-1$ and $P_2(a)=a(a-1)/2$. Assuming 
they are valid for any real $a$ we get at once
\eq
\sum_{n>0}'n^k=2^k P_{k+1}(\um)/(2^{k+1}-1)~,
\label{sumk}
\en
which turns out to coincide with the value of the Riemann zeta function 
$\zeta(-k)$. In this spirit we get the identities
\ea
&&\prod_{n>0}'c=\exp[\log c~\sum_{n>0}'n^0]=\frac1{\sqrt{c}}\\
&&\prod_{n\in{\N+\um}}'c=\exp[\log c~\sum_{n>0}'(n-\um)^0]=1
\label{resultb}\\  
&&\prod_{n>0}'n=\sqrt{2\pi}~~,~\prod_{n>0}'(n-\um)=\sqrt{2}~
\label{resultc}
\na
Which can directly applied to Eq.(\ref{lapla}), yielding
\eq
\vert-\partial^2\vert=\prod_{m>0}'\prod_{n>0}
\left[1+\frac{r^2}{t^2}\left(\frac{m-\um}{n}
\right)^2\right]~.
\en
Using the known identity
\eq
\prod_{n>0}(1+\frac{\alpha^2}{n^2})=\frac{\sin i\pi\alpha}{i\pi\alpha}=
\frac{e^{-i\pi\alpha}}{2\pi\alpha}(1-e^{i2\pi\alpha})~,
\en
combined with Eq.s (\ref{suma}),(\ref{resultb}) and (\ref{resultc}) 
leads us finally to the finite result
\eq
\vert-\partial^2\vert=\frac1{\sqrt{2}}q^{-\frac1{48}}\prod_{m>0}(1-q^{m-\um})
,~q=e^{i2\pi\frac rt}
\en
which can be written in a number of equivalent ways:
\eq
\vert-\partial^2\vert\equiv\sqrt{\frac{\theta_4(\tau)}{2\eta(\tau)}}
\equiv\frac{\eta(\tau/2)}{\sqrt{2}\,\eta(\tau)}\equiv
\frac{\eta(-2/\tau)}{\eta(-1/\tau)}~,
\en
where $\tau=i\frac rt$, $\theta_4$ is a Jacobi $\theta$-function and 
$\eta(\tau)=q^{\frac1{24}}\prod_{n>0}(1-q^n)$ is the Dedekind's eta function.

In conclusion, the first term of the asymptotic expansion (\ref{genera}) 
applied to the U-shaped Wilson line can be written in the form
\ea
\bra\U(r,t)\ket\propto F(t/r)\,e^{-rt\,\sigma-(2t+r)\,\lambda-r\R_b\sigma/2}
+\,
\cdot\cdot\cdot
\label{uushape}\\
F(t/r)=\left(\frac{\eta(i\frac tr)}{\eta(i\frac{2t}r)}\right)^{\frac{d-2}2} ~.
\na
As a check, it is easy to see that the static potential defined as 
$V(r)=-\lim_{t\to\infty}\log(\bra\U(r,t)\ket)$ contains the $1/r$ L\"uscher term
with the right coefficient.
 
Observing these  effects is very challenging.
Strange though it may seem, boundary terms constitute the major obstruction 
to detect universal string signals.
 
Indeed finite size effects in interfaces, where the involved string world-sheet 
has no boundary, is the first place where these 
effects were clearly seen \cite{Caselle:1992ue}.    

Similarly, in the Polyakov loop correlators the boundaries contribution does not 
depend on their distance; indeed in such a system are concentrated the main theoretical 
efforts to understand the physics of the confining string, after the introduction of 
the variance reduction method \cite{Luscher:2001up}.  For a recent discussion 
on this argument and a complete list of references see \cite{Caselle:2005vq}.

In the case of Wilson loops the boundary term increases with the loop size, hence it 
is much more difficult to achieve accurate results. As a matter of fact only in very 
simple systems, like  in 3D $\Z_2$ gauge model \cite{Caselle:1996ii} or even in 
percolation processes \cite{Gliozzi:2005ny} high precision estimates have been reached.

In the case of open Wilson lines the situation dramatically worsen,
the reason being that there are two competing requirements: the shape effects 
would be  visible in sufficiently large open Wilson lines provided the string does 
not break, hence one has to chose a region where $\R_b$ is large, however 
this implies, as  Eq.(\ref{uushape}) shows, an exponential weakening of the 
signal, because of the contribution of the free boundary.      
Studies are under way to gain more insight into these operators.

\end{document}